\documentclass[twoside,9pt]{article}
\usepackage{rsc}
\hypersetup{colorlinks=true, allcolors=blue}

\usepackage{amsmath}
\usepackage{tikz}
\usetikzlibrary{arrows.meta, positioning, shapes.geometric}
\begin{document}
\nocite{rsc-control}
\begin{flushleft}
{\fontsize{18pt}{20pt}\selectfont\textbf{Bubble detachment from circular cavities and flat surfaces}}
\end{flushleft}
{\fontsize{12pt}{14pt}\selectfont Ianto Cannon\textsuperscript{a\dag}, Stefan Endres\textsuperscript{b,c}, Lutz M\"adler\textsuperscript{b,c} and Marc Avila\textsuperscript{a,c}}\par\medskip
Received 00th January 20xx, Accepted 00th January 20xx
DOI: 10.1039/x0xx00000x\vspace{0.5cm}

We determine the maximum stable volume of bubbles attached to flat surfaces as a function of contact radius, contact angle, and capillary length. By solving the Young--Laplace equation using a shooting method, we calculate equilibrium bubble shapes and determine when bubbles detach through three distinct modes: necking at the cavity, sideways instability, and spreading followed by necking on the surrounding surface. Our predicted detachment volumes agree well with experimental measurements of electrolytic, boiling, and carbonated bubbles, as well as pendant drops. These results bring together and extend previous analytical estimates and, by symmetry, also apply directly to pendant drops.
\vspace{1cm}

\section{Introduction}

\subsection{Motivation}

Annually, around $10^{11}$\,kg of hydrogen gas is used in industry, mostly to produce fertilisers via the Haber-Bosch process\cite{bosch32development} and to reduce metal ores in steelmaking and aluminium production\cite{kullmann23role}. Currently, more than 95\% of hydrogen is produced from fossil fuels, largely from natural gas via steam methane reforming\cite{franchi20hydrogen}, which releases significant quantities of CO$_2$. Diversifying away from fossil fuels is essential to maintain a stable Earth climate system and ensure the long-term survival of human society\cite{steffen18trajectories}. Water electrolysis driven by renewable electricity is a promising alternative\cite{kullmann23role}, splitting water into hydrogen and oxygen without direct CO$_2$ emissions. However, gas bubbles that form and remain attached to the electrode surface limit the efficiency of water electrolysis. These bubbles block the active area, impede ion transport, and increase ohmic resistance\cite{vogt89problem,penas19decoupling,angulo20influence,ross25impact,ross26transient}. Electrolysis will also be important for life-support systems on future manned lunar and Martian bases, where water extracted from local ice deposits could provide a source of hydrogen and oxygen\cite{matsushima06water}. Understanding and controlling bubble detachment is therefore relevant both to improving electrolysis efficiency on Earth and to enabling efficient resource utilisation in low-gravity environments\cite{brinkert22fundamentals}.

Bubble detachment also plays a role in the transport of naturally occurring gases. Methane seepage from the seafloor releases an estimated $6$--$12$\,Mt of methane per year from anoxic sediments, geological reservoirs, and degrading methane-hydrate deposits\cite{weber19global}. Bubbles with radii below approximately 3\,mm can dissolve and biodegrade in the water column before reaching the ocean surface\cite{mcginnis06fate}, whereas larger bubbles can escape into the atmosphere. Since methane is a potent greenhouse gas, understanding the conditions governing bubble detachment and subsequent transport is relevant to quantifying its environmental impact.

\subsection{Background}

\begin{figure}[!b]
\includegraphics[width=\textwidth]{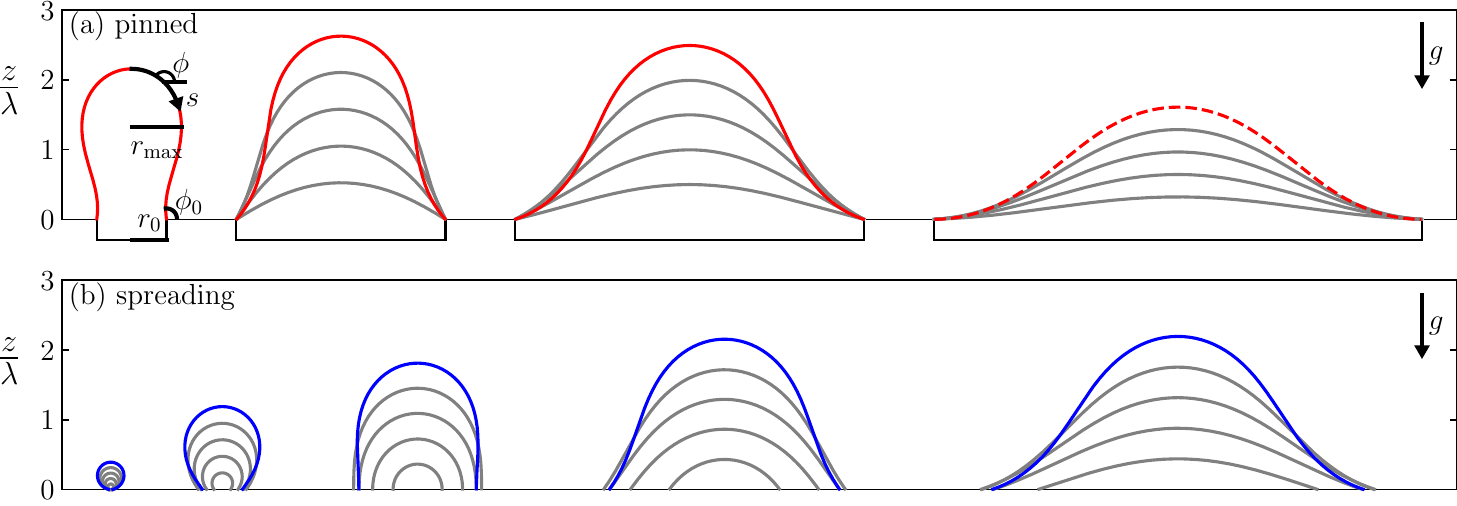}
\caption{(a)~Profiles of pinned bubbles with contact radius $r_0/\lambda\in{0.5,1.5,2.5,3.5}$, from left to right. The largest pinned bubble for each contact radius is shown in red. The $r_0=3.5\lambda$ profile is dashed to indicate the limiting case $\phi_0=\pi$, for which the bubble detaches via a non-axisymmetric instability. The coordinate system $(\phi,s)$ used to solve Equation~\ref{eqShape} and the bubble properties $\phi_0$, $r_0$, and $r_\mathrm{max}$ are shown in black. (b)~Profiles of spreading bubbles with contact angle $\phi_0/\pi\in{0.1,0.3,0.5,0.9}$, from left to right. The largest spreading bubble for each contact angle is shown in blue.}
\label{figProfile}
\end{figure}

For a bubble attached to a horizontal surface (illustrated in Figure~\ref{figProfile}) the scale over which surface tension $\sigma$ balances fluid density contrast $\rho_\mathrm{out} - \rho_\mathrm{in}$ and gravity $g$ is the capillary length\cite{laplace98traite},
\begin{equation}
\lambda \equiv \sqrt\frac{\sigma}{(\rho_\mathrm{out} - \rho_\mathrm{in})g}.
\end{equation}
The capillary length sets the radius of curvature of the bubble interface due to gravity. Two distinct attachment modes are recognised in the literature\cite{chesters78modes}. \emph{Pinned} bubbles have a fixed contact radius $r_0$, with contact angle $\phi_0$ changing as the bubble inflates. \emph{Spreading} bubbles have a fixed contact angle $\phi_0$, with contact radius $r_0$ changing as the bubble inflates.

For \emph{pinned} bubbles, the classical Tate volume\cite{tate64magnitude} assumes that a bubble detaches when the contact angle reaches $\phi_0=\pi/2$, balancing buoyancy and surface tension to give 
\begin{equation}\label{eqTate}
V=2\pi\lambda^2 r_0.
\end{equation} 
Chesters\cite{chesters77analytical} refined this in the limit $r_0/\lambda\to0$, also assuming detachment at $\phi_0=\pi/2$. Subsequent experiments, however, show systematic deviations from both models\cite{gerlach05quasistatic,lesage13experimental,mori01bubble,sasetty23stability,gunde01measurement}, due to non-spherical interfaces and the onset of asymmetrical instabilities. Stability analyses indicate that the bubble becomes unstable to asymmetrical perturbations for cavity radii $r_0>3.219\,\lambda$\cite{michael76equilibrium}, and no stable bubble exists at fixed volume for $r_0>3.832\,\lambda$\cite{plateau73statique}. For such large cavities, lateral detachment from the rim is observed experimentally\cite{sasetty23stability}.

The study of \emph{spreading} bubble departure dates back to Fritz\cite{fritz35berechnung}, who numerically estimated the maximum bubble volume as 
\begin{equation}\label{eqFritz}
V=0.887\lambda^3\phi_0^3
\end{equation}
in the limit of small contact angles $\phi_0$. Focusing instead on large contact angles, Phan et al.\cite{phan09surface} developed an empirical model for bubble departure from nanocoated surfaces. More recent models account for contact-angle hysteresis\cite{degennes85wetting}, using advancing, receding\cite{allred21role,demirkir24life}, or dynamic\cite{mukherjee07numerical} angles as inputs. Experiments, however, reveal systematic deviations from these predictions, particularly for bubbles on superhydrophobic and high-contact-angle surfaces\cite{huang26effects}.

Bubble detachment can also be triggered by coalescence with a neighbouring bubble. Whether neighbouring bubbles coalesce is determined by their width relative to the distance between nucleation sites\cite{bonjour00experimental,raman25electrolytic}. Following coalescence, the merged bubble may either remain attached or detach from the surface, depending on the balance between adhesion and viscous dissipation\cite{demirkir25jump}.

\subsection{Aims}

Here, we treat pinned and spreading bubbles together, rather than as separate problems. We address three fundamental questions: 

\textbf{Q1} What is the volume of the bubble when it detaches?

\textbf{Q2} How wide does the bubble grow before it detaches?

\textbf{Q3} How does it detach?

We consider a simplified system consisting of a circular cavity of radius $r_0$ containing an already-nucleated bubble, surrounded by a smooth horizontal surface with a single-valued, non-hysteretic contact angle $\phi_0$. That is, the contact angle is independent of whether the three-phase contact line advances or recedes. The fluid is quiescent, so forces are purely hydrostatic, and the system is physically equivalent to a pendant drop. Throughout this paper, we refer to it as a bubble, but our results apply equally to pendant drops. We determine the detachment size for pinned and spreading bubbles across the full range of cavity sizes, capillary lengths ($0< r_0/\lambda<\infty$), and contact angles ($0<\phi_0<\pi$).

The remainder of this paper is organised as follows. In Section~\ref{secMethods}, we outline the numerical methods used to solve the governing Young--Laplace equation. In Section~\ref{secResults}, we present the results for pinned and spreading bubbles separately, examining their detachment criteria (Section~\ref{secGrowth}), maximum stable volumes (Section~\ref{secVol}), coalescence dynamics (Section~\ref{secCoal}), and nucleation characteristics (Section~\ref{secNuc}). Finally, in Section~\ref{secUnpin} we connect these two modes to establish a unified framework for bubble evolution, before drawing our conclusions in Section~\ref{secConclusions}.
\section{Methods}\label{secMethods}

\subsection{Governing equation}

To calculate the bubble shape, we use and extend the method developed by\citet{fritz35berechnung}. Consider a bubble attached to a solid horizontal surface, as depicted in Figure~\ref{figProfile}. The bubble interface satisfies the Young--Laplace equation\cite{degennes04capillarity}:
\begin{equation}\label{eqLaplace}
\sigma \left( \frac{1}{R_1} + \frac{1}{R_2} \right) = p_\mathrm{in} - p_\mathrm{out},
\end{equation}
where $R_1$ and $R_2$ are the principal radii of curvature at a point on the interface, and $p_\mathrm{in}$ and $p_\mathrm{out}$ are the pressures inside and outside the bubble, respectively. Assuming both fluid phases are in hydrostatic equilibrium, the pressures at height $z$ are given in terms of the reference pressures $p_{0\mathrm{in}}$ and $p_{0\mathrm{out}}$ at the substrate ($z=0$):
\begin{align}
p_\mathrm{in}(z)  &= p_\mathrm{in0}  - \rho_\mathrm{in} g z, \\
p_\mathrm{out}(z) &= p_\mathrm{out0} - \rho_\mathrm{out} g z.
\end{align}
Substituting into Equation~\ref{eqLaplace} and dividing through by the surface tension gives:
\begin{equation}\label{eqPressure}
  \frac{1}{R_1} + \frac{1}{R_2} = \frac{p_\mathrm{in0} - p_\mathrm{out0}}{\sigma} + \frac{z}{\lambda^2}.
\end{equation}
To eliminate the pressure difference $p_\mathrm{in0} - p_\mathrm{out0}$, we evaluate Equation~\ref{eqPressure} at the apex of the bubble, where $z = h$ and the interface is spherical, so we let $R_1 = R_2 = R_h$, the apex radius of curvature:
\begin{equation}
\frac{2}{R_h} = \frac{p_\mathrm{in0} - p_\mathrm{out0}}{\sigma} + \frac{h}{\lambda^2}.
\end{equation}
Subtracting this from Equation~\ref{eqPressure} gives:
\begin{equation}\label{eqPressCurv}
\frac{1}{R_1} + \frac{1}{R_2} = \frac{2}{R_h} + \frac{z - h}{\lambda^2}.
\end{equation}
Assuming the bubble is axisymmetric about the vertical axis, we describe the interface shape in terms of arc length $s$ from the apex, distance from the axis $r$, and the angle $\phi$ between the interface tangent and the horizontal. The two principal radii of curvature can be found through trigonometry\cite{bashforth83attempt}:
\begin{equation}
R_1 = \frac{ds}{d\phi}, \qquad R_2 = \frac{r}{\sin\phi}.
\end{equation}
Substituting into Equation~\ref{eqPressCurv} and multiplying by the capillary length yields the governing equation for the bubble shape:
\begin{equation}\label{eqShape}
\lambda\frac{d\phi}{ds} + \frac{\lambda\sin\phi}{r} = \frac{2\lambda}{R_h} + \frac{z - h}{\lambda}.
\end{equation}
This equation describes the curvature of an axisymmetric interface under gravity. Here, all lengths are scaled by the capillary length $\lambda$, so that Equation~\ref{eqShape} is dimensionless, implying self-similar bubble shapes and a universal scaling with $\lambda$.

\subsection{Numerical solution}

We solve Equation~\ref{eqShape} using a shooting method, initiating the integration at the apex ($r=0$, $z=h$, $\phi=\pi$) and sampling $10^4$ logarithmically spaced initial guesses for $R_h$ over $10^{-3}\lambda<R_h<10^3\lambda$. The equations are integrated along the arc length $s$ with a step size of $10^{-5}\lambda$ using the Adams--Bashforth method\cite{bashforth83attempt}. The complete set of calculations requires approximately 250\,s on a single core of an M4 MacBook Pro. The C code and accompanying data are freely available\cite{cannon26bubble}.

For pinned bubbles, the boundary condition $r=r_0$ is imposed at the substrate ($z=0$). For a given cavity radius $r_0$, this yields a family of equilibrium profiles with different apex radii $R_h$, shown in Figure~\ref{figProfile}a. For spreading bubbles, the boundary condition $\phi=\phi_0$ is imposed at the substrate ($z=0$). The resulting profiles are shown in Figure~\ref{figProfile}b for a sequence of contact angles and apex radii $R_h$.

\section{Results and Discussion}\label{secResults}

\subsection{Bubble detachment criteria}\label{secGrowth}

\begin{figure}
\includegraphics[width=\textwidth]{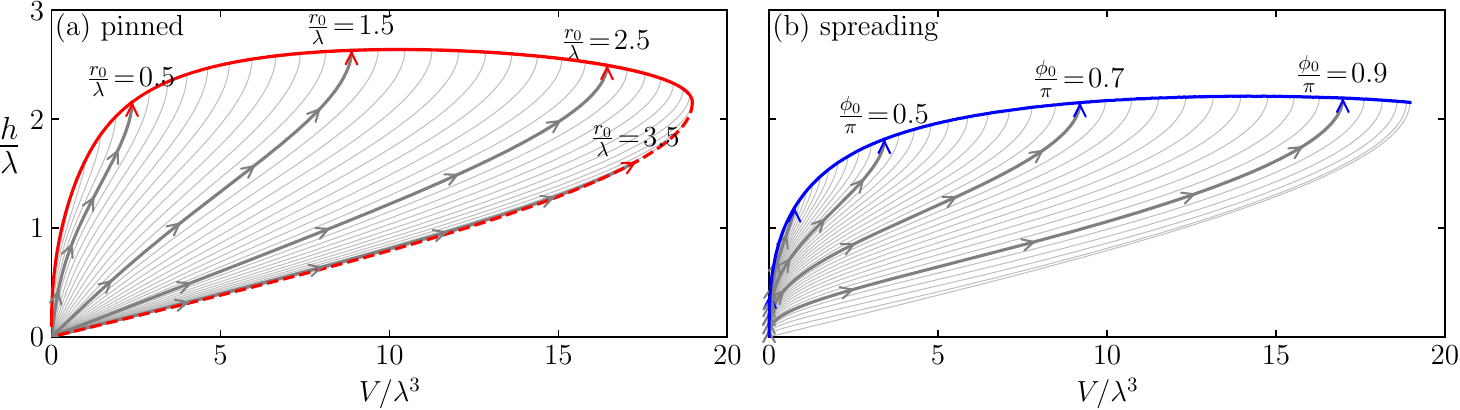}
\caption{Bubble height $h$ and volume $V$ during evolution. (a)~Grey lines: pinned bubbles with constant contact radius $r_0$. Solid red line: locus where $\partial V/\partial h=0$, and the bubbles detach axisymmetrically. Dashed red line: locus where $\phi_0=\pi$ and the bubbles detach from the side of the cavity. (b)~Grey lines: spreading bubbles with constant contact angle $\phi_0$. Solid blue line: locus where $\partial V/\partial h=0$, and the bubbles detach axisymmetrically. In (a)~and (b), thick grey lines and arrows correspond to the bubbles shown in Figure~\ref{figProfile}.}
\label{figHeight}
\end{figure}

To characterise the evolution of a growing pinned bubble, we plot the bubble height $h$ as a function of volume $V$ in Figure~\ref{figHeight}a, where the bubble volume
\begin{equation}
V=\int_0^h\pi r^2\,\mathrm{d}z,
\end{equation}
is integrated using the midpoint rule. Each family of solutions in Figure~\ref{figProfile}a gives rise to a curve $h(V)$ in Figure~\ref{figHeight}a. Upon nucleation inside the cavity, the bubble has zero volume and zero height, corresponding to the bottom-left point in Figure~\ref{figHeight}a. As the bubble grows, its volume $V$ and height $h$ increase, and it remains attached to the substrate until one of two conditions is met:
\begin{enumerate}
\item The bubble volume reaches a point where the $h(V)$ curve becomes vertical (i.e., $\partial V/\partial h = 0$), meaning the height can increase without a further increase in volume. The solution is no longer unique, and the interface becomes unstable, leading to axisymmetric detachment.
\item The contact angle $\phi_0$ reaches $\pi$, at which point the interface is horizontal at the contact line and becomes unstable to non-axisymmetric perturbations\cite{michael76equilibrium}. The bubble shifts to one side of the cavity and detaches laterally. This occurs for all $h(V)$ curves with $r_0>3.219\,\lambda$, at the lower right of Figure~\ref{figHeight}a.
\end{enumerate}
We truncate each of the $h(V)$ curves in Figure~\ref{figHeight}a when either of these conditions is met. We note that the tallest possible bubble occurs in a cavity of radius $r_0=1.720\,\lambda$ and has height $h=2.642\,\lambda$, which is considerably less than the analytical limit of $h=3.42\,\lambda$ established for two-dimensional columnar pendant drops\cite{sumesh10possible}.

In Figure~\ref{figHeight}b, we perform the same $h(V)$ analysis for spreading bubbles. The figure shows the possible volumes and heights for contact angles spanning the full range $0<\phi_0<\pi$. As for pinned bubbles, each curve reaches a point where $h(V)$ becomes vertical ($\partial V/\partial h=0$). This point defines the detachment volume, and the curve is truncated there. In the spreading bubble case, detachment is always axisymmetric, and the tallest spreading bubble is shorter, with $h=2.213\,\lambda$ and contact angle $\phi_0=0.8330\pi$.

\subsection{Bubble volume}\label{secVol}

\begin{figure}\centering
\includegraphics[width=\columnwidth]{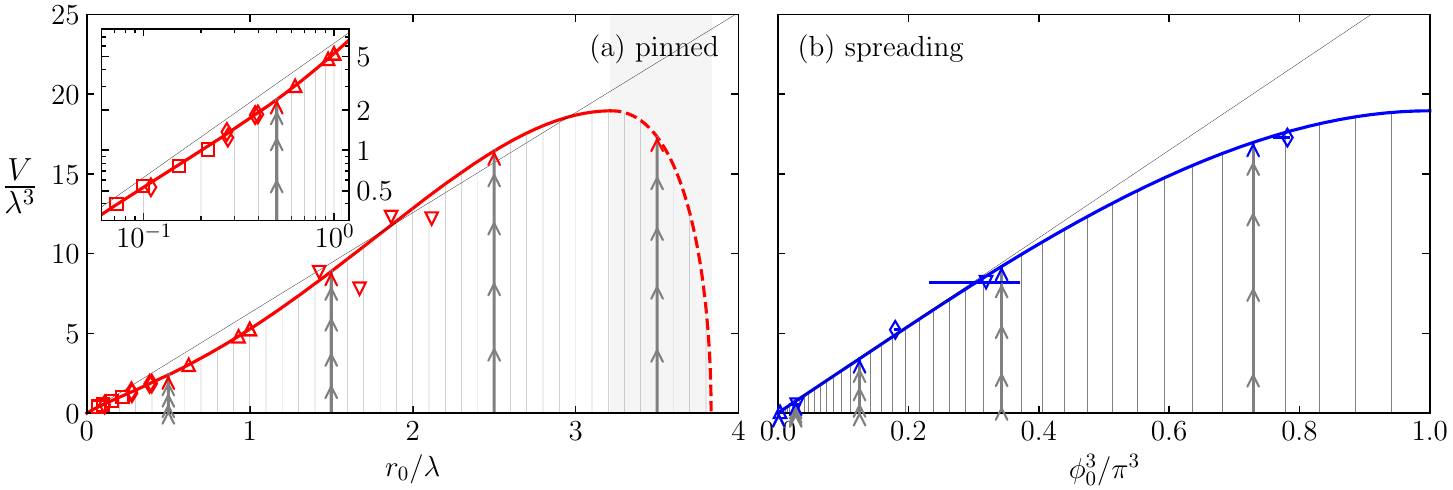}
\caption{(a)~Volume $V$ of pinned bubbles. The red line shows the detachment volume as a function of contact radius $r_0$, while the thin black line, $V=2\pi\lambda^2 r_0$, shows the Tate prediction\cite{tate64magnitude}. The shaded region $3.219\,\lambda < r_0 < 3.832\,\lambda$ and dashed red line indicate where $\phi_0$ reaches $\pi$, and the bubble becomes unstable to asymmetric perturbations. Empty symbols denote detachment volumes measured in experiments: squares\cite{lesage13experimental}, diamonds\cite{mori01bubble}, down-triangles\cite{sasetty23stability}, and up-triangles\cite{gunde01measurement}. Inset: same data on a logarithmic scale, showing the small-radius range in greater detail. (b)~ volume $V$ of spreading bubbles. The solid blue line shows the detachment volume as a function of contact angle $\phi_0$, which is cubed to allow comparison with the Fritz\cite{fritz35berechnung} prediction $V=0.887\lambda^3\phi_0^3$ (thin black line). Empty symbols denote detachment volumes measured in experiments: up-triangles\cite{demirkir24life}, down-triangles\cite{allred21role}, and diamonds\cite{huang26effects}; error bars indicate advancing and receding contact angles. In (a)~and (b), thick grey lines and arrows correspond to the bubbles shown in Figure~\ref{figProfile}.}
\label{figVol}
\end{figure}

We take the detachment volume to be the entire bubble volume above the contact line at the moment of detachment. We note, however, that some gas may remain attached to the surface after detachment, so the true released volume may be slightly smaller.

Figure~\ref{figVol}a shows the detachment volume of pinned bubbles. It initially increases with contact radius, closely following the Tate prediction\cite{tate64magnitude}; deviations from this prediction arise from the non-spherical interface shape induced by buoyancy. These deviations are consistent with experimental measurements of bubbles emitted from tubes in water\cite{lesage13experimental}, bubbles in carbonated water from millimetric cavities\cite{mori01bubble}, and pendant drops\cite{gunde01measurement}. The largest pinned bubble occurs at $r_0=3.219\,\lambda$, with a volume of $V=18.96\,\lambda^3$. Beyond this point, for $r_0>3.219\,\lambda$, $\phi_0$ reaches $\pi$ and the interface becomes unstable to non-axisymmetric perturbations\cite{michael76equilibrium}, causing the bubble to slide to one side and detach laterally; the detachment volume then decreases with increasing contact radius. No stable solution exists for $r_0>3.832\,\lambda$\cite{plateau73statique}: above this radius, Equation~\ref{eqShape} has no solution with $\phi_0<\pi$, and lateral detachment occurs immediately, consistent with the dripping behaviour reported in sessile drop experiments\cite{sasetty23stability}.

Figure~\ref{figVol}b shows the detachment volumes of spreading bubbles. For small contact angles, the Fritz prediction is accurate. As $\phi_0\to\pi$, however, the detachment volume saturates below Fritz's prediction. Our results are confirmed by measurements of water electrolysis\cite{demirkir24life}, boiling\cite{allred21role}, and air bubbles over superhydrophobic surfaces\cite{huang26effects}. The largest spreading bubble has $V=18.96\,\lambda^3$, and is identical in shape to the largest pinned bubble. 

\subsection{Bubble coalescence}\label{secCoal}

\begin{figure}\centering
\includegraphics[width=\columnwidth]{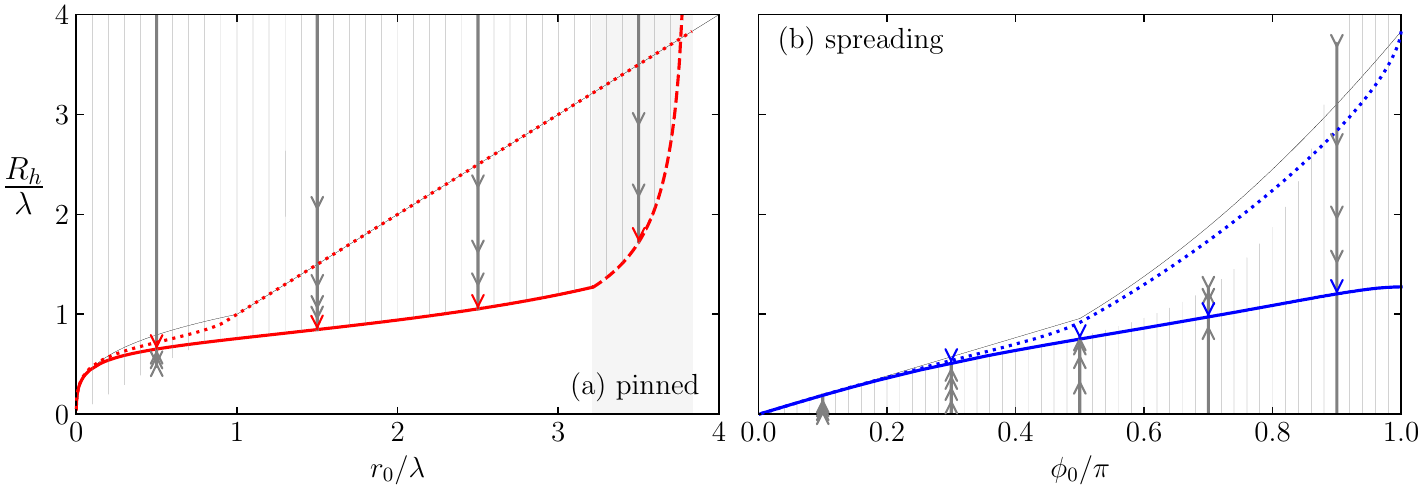}
\caption{(a)~Apex radius of curvature $R_h$ of pinned bubbles as a function of contact radius $r_0$. The red line shows $R_h$ at detachment. The shaded region $3.219\,\lambda<r_0<3.832\,\lambda$ and dashed red line indicate where $\phi_0$ reaches $\pi$, and the bubble becomes unstable to asymmetric perturbations. The dotted red line shows the maximum horizontal radius $r_\mathrm{max}$ reached during evolution, and the thin black line shows the piecewise fit $r_\mathrm{max}/\lambda=(r_0/\lambda)^{1/3}$ for $r_0<\lambda$ and $r_\mathrm{max}=r_0$ for $r_0>\lambda$. (b)~Apex radius of curvature $R_h$ of spreading bubbles as a function of contact angle $\phi_0$. The solid blue line shows $R_h$ at detachment; the dotted blue line shows the maximum horizontal radius $r_\mathrm{max}$ reached during evolution; and the thin black line shows the piecewise fit $r_\mathrm{max}/\lambda=3.832\phi_0/(2\pi)$ for $\phi_0<\pi/2$ and $r_\mathrm{max}/\lambda=3.832(\phi_0/\pi)^2$ for $\phi_0>\pi/2$. In (a)~and (b), vertical grey solid lines show the evolution of $R_h$ up to detachment, with arrowheads corresponding to the bubbles shown in Figure~\ref{figProfile}.}
\label{figRad}
\end{figure}

When two bubbles nucleate near each other on a surface, they may coalesce, and the energy released during coalescence can lead to detachment\cite{demirkir25jump}. Assuming that coalescence takes place when neighbouring bubbles touch, and that the bubbles have the same shape, coalescence occurs when the distance between their nucleation centres is less than or equal to $2r_\mathrm{max}$\cite{raman25electrolytic}, where $r_\mathrm{max}$ is the maximum horizontal radius of the bubble, as labelled in Figure~\ref{figProfile}. Figure~\ref{figRad}a shows the maximum value of $r_\mathrm{max}$ reached by pinned bubbles during their evolution up to detachment. The maximum horizontal radius follows the piecewise fit:
\begin{alignat}{2}
r_\mathrm{max} &\approx r_0^{1/3}\lambda^{2/3} \quad &&\text{for } r_0 \le \lambda, \\
r_\mathrm{max} &= r_0 \quad &&\text{for } r_0 > \lambda.
\end{alignat}
For small contact radii, this means that bubbles are expected to coalesce when the distance between their nucleation centres is less than $2r_0^{1/3}\lambda^{2/3}$. For large contact radii, $r_\mathrm{max}=r_0$ indicates that the bubble does not protrude beyond the cavity. Hence, coalescence cannot occur between bubbles nucleated on neighbouring cavities with radii greater than $\lambda$.

Figure~\ref{figRad}b shows that the maximum value of $r_\mathrm{max}$ reached during spreading bubble evolution is also a piecewise function:
\begin{alignat}{2}
r_\mathrm{max} &\approx 3.832\,\lambda\phi_0/(2\pi) \quad &&\text{for } \phi_0 \le \pi/2, \\
r_\mathrm{max} &\approx 3.832\,\lambda\left(\phi_0/\pi\right)^2 \quad &&\text{for } \phi_0 > \pi/2.
\end{alignat}
Thus, the critical separation scales linearly with $\phi_0$ for small contact angles, but with $\phi_0^2$ for larger ones. This means nucleation sites must be packed closely on hydrophilic surfaces to induce coalescence, whereas bubbles spread wider on hydrophobic surfaces (large $\phi_0$), and hence coalescence-induced detachment can be triggered with a significantly larger pitch between cavities.

\subsection{Bubble nucleation}\label{secNuc}

Figure~\ref{figRad} also shows the evolution of the bubble apex radius of curvature $R_h$. Pinned bubbles nucleate with an infinite radius of curvature, corresponding to a flat interface spanning the cavity and, as follows from the Young--Laplace equation~(\ref{eqLaplace}), zero pressure difference across the interface. In contrast, spreading bubbles nucleate with $R_h=0$, corresponding to an infinitesimal spherical cap and a diverging pressure difference. This large pressure difference helps explain why nucleation of spreading bubbles is energetically unfavourable. Instead, nucleation generally occurs as a pinned bubble within a microscopic cavity.

\subsection{Unpinning of the contact line}\label{secUnpin}

\begin{figure}\centering
\includegraphics[width=\columnwidth]{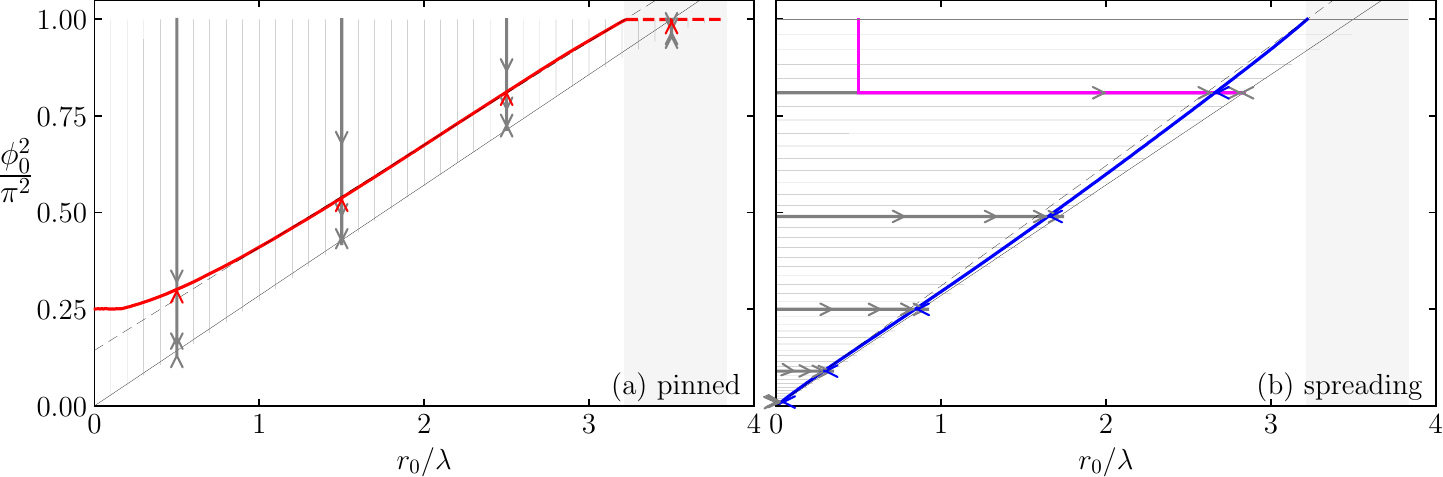}
\caption{(a)~Contact angle $\phi_0$ of pinned bubbles with fixed contact radius $r_0$. The red line shows $\phi_0$ at detachment and is dashed where $\phi_0=\pi$ and detachment occurs non-axisymmetrically. The dashed black line shows the fit $(\phi_0/\pi)^2=0.2661r_0/\lambda+0.1436$. (b)~Contact radius $r_0$ of spreading bubbles with fixed contact angle $\phi_0$. The magenta line is the complete bubble evolution in a system with cavity radius $r_c=0.5\,\lambda$ and wetting angle $\phi_w=0.9\pi$. The blue line shows $r_0$ at detachment, with the dashed black line showing the fit $r_0=3.219\,\lambda(\phi_0/\pi)^2$. In (a)~and (b), grey lines and arrows show the evolution up to detachment, the thin solid line $r_0=3.5\,\lambda(\phi_0/\pi)^2$ is a fit to the turning points, and the shaded region indicates $3.219\,\lambda<r_0<3.832\,\lambda$.}
\label{figAng}
\end{figure} 

Figure~\ref{figAng}a tracks how the contact angle of pinned bubbles evolves up to detachment. Every pinned bubble starts as a flat interface, so $\phi_0=\pi$ initially. As the bubble inflates, the contact angle decreases until it reaches a minimum, coinciding with the appearance of a buoyancy-induced inflection point near the bubble's foot (visible in Figure~\ref{figProfile}), which then causes the contact angle to increase again before detachment. When the contact radius is small ($r_0<0.1\,\lambda$), the contact angle at detachment is close to $\pi/2$, consistent with the assumptions made by Tate and Chesters\cite{tate64magnitude,chesters77analytical}. For larger contact radii, however, $\phi_0$ at detachment exceeds $\pi/2$. This is why the Tate and Chesters predictions agree well with our results and experiments of others\cite{lesage13experimental,mori01bubble,sasetty23stability,gunde01measurement} for small contact radii but diverge as buoyancy effects grow stronger.

Figure~\ref{figAng}b tracks how the contact radius $r_0$ of spreading bubbles evolves up to detachment. These bubbles start with $r_0=0$, and the contact radius grows as they inflate. As with pinned bubbles, buoyancy induces an inflection point near the bubble's foot, causing the contact radius to reach a maximum before the contact line recedes and the bubble detaches. In the $(r_0,\phi_0)$ plane, the locus of maximum contact radii for spreading bubbles coincides with the locus of minimum contact angles for pinned bubbles, and both are well described by the fit $\phi_0/\pi\approx\sqrt{r_0/(3.5\,\lambda)}$. At detachment, the contact radius scales roughly as the square of the contact angle, following $r_0=3.219\,\lambda(\phi_0/\pi)^2$.

\begin{figure}[h]
\centering
\begin{tikzpicture}[
    block/.style={draw, rectangle, rounded corners, align=center, minimum width=2.5cm, minimum height=0.8cm},
    decis/.style={draw, diamond, aspect=2, align=center, inner sep=0pt, minimum width=2.5cm, minimum height=0.8cm},
    red/.style={draw=red},
    blu/.style={draw=blue},
    arrow/.style={->}
]
\pgfmathsetmacro{\colA}{0}
\pgfmathsetmacro{\colB}{3.5}
\pgfmathsetmacro{\colC}{7}
\pgfmathsetmacro{\colD}{10.5}
\pgfmathsetmacro{\rowTop}{2}    
\pgfmathsetmacro{\rowUp}{1}      
\pgfmathsetmacro{\rowMid}{0}     
\pgfmathsetmacro{\rowBot}{-1}   
\node[block, red] (pinn) at (\colA, \rowMid) {pinned bubble\\ with $r_0=r_c$};
\node[decis, red] (dec1) at (\colB, \rowMid) {is $\frac{r_c}{\lambda}\!<\!3.5\frac{\phi_w^2}{\pi^2}$?};
\node[decis, red] (dec2) at (\colC, \rowUp) {is $r_c\!>\!3.219\,\lambda$?};
\node[block, blu] (spre) at (\colC, \rowBot) {spreading bubble\\ with $\phi_0=\phi_w$};
\node[block, blu] (detB) at (\colD, \rowBot) {detach from\\ surface};
\node[block, red, fill=gray!20, dashed] (detS) at (\colD, \rowMid) {detach from\\ side of cavity};
\node[block, red] (detA) at (\colD, \rowTop) {detach from\\ centre of cavity};
\draw[arrow, red] (pinn.east) -- (dec1.west);
\draw[arrow, red] (dec1) |- node[above, near end] {No} (dec2.west);
\draw[arrow, blu] (dec1) |- node[above, near end] {Yes} (spre.west);
\draw[arrow, blu] (spre) -- (detB.west);
\draw[arrow, red] (dec2) |- node[above, near end] {No} (detA.west);
\draw[arrow, red] (dec2) |- node[above, near end] {Yes} (detS.west);
\end{tikzpicture}
\caption{Decision tree for bubble growth. A bubble nucleates in a circular cavity of radius $r_c$, surrounded by a flat substrate. The three-phase wetting angle on the substrate is $\phi_w$. Red branches correspond to pinned bubbles and blue branches to spreading bubbles.}
\label{figTree}
\end{figure}
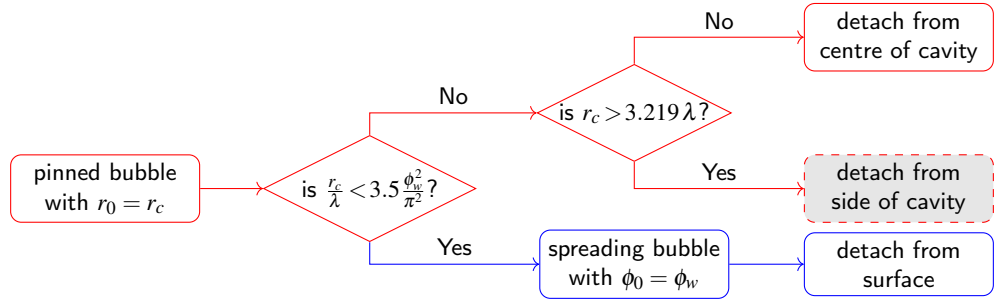

Up to this point in the discussion, we have considered the results for pinned and spreading bubbles separately. Bubbles typically nucleate in the pinned mode\cite{chi-yeh65mechanism} with $r_0=r_c$, the radius of the cavity. However, the contact line will become unpinned if the contact angle $\phi_0$ reaches the wetting angle $\phi_w$ of the three-phase system\cite{chesters78modes}, it will then spread with constant contact angle $\phi_0=\phi_w$. An example case $r_c=0.5\,\lambda, \phi_w=0.9\pi$ is illustrated in Figure~\ref{figAng}b; the bubble nucleates pinned, and during evolution it becomes unpinned, and its contact line spreads across the surface surrounding the cavity before it reaches the maximum radius, recedes a little, and finally detaches. In general, we can see that the unpinning transition will occur if the crossover between the pinned mode $r_0=r_c$ and the spreading mode $\phi_0=\phi_w$ is reached before the minimum contact angle is reached, i.e., if the crossover point is above the line of best fit:
\begin{equation}
r_c < 3.5\,\lambda\left(\frac{\phi_w}{\pi}\right)^2.
\end{equation}
Otherwise, the wetting angle will not be reached, and the bubble will remain in the pinned mode until detachment. Detachment from the pinned mode occurs axisymmetrically if the cavity radius $r_c<3.219\,\lambda$, and non-axisymmetrically for large cavities $r_c\ge3.219\,\lambda$. Figure~\ref{figTree} summarises these results in a decision tree for bubble evolution up to detachment. Given $\phi_0$, $r_0$, and the capillary length $\lambda$ of a specific system, the detachment size is determined by following this tree, using the functions plotted in Figure~\ref{figVol}. Table~\ref{tabSig} gives the properties of the limiting bubble configurations, where we see that the largest pinned and spreading bubbles in fact correspond to the same solution of Equation~\ref{eqShape}.

\begin{table}\small
\caption{Key properties of significant bubbles. Shown are the contact angle $\phi_0$, contact radius $r_0$, apex height $h$, apex radius of curvature $R_h$, and volume $V$, expressed in units of the capillary length $\lambda$. The corresponding radius $r_\mathrm{det}$ of the detached bubble is given for water at standard temperature and pressure, where $\lambda=2.793$\,mm. Extremal values are typeset in boldface.}\label{tabSig}
\begin{center}
\begin{tabular}{lcccccc}
\hline
Significance              & $\phi_0/\pi$          & $r_0/\lambda$& $h/\lambda$ & $R_h/\lambda$ & $V/\lambda^3$ & $r_\mathrm{det}$/mm \\ \hline
largest pinned bubble     & 1.000  & 3.219        & 2.152        & 1.274       &\textbf{18.96} & \textbf{4.620}\\
largest spreading bubble  & 1.000  & 3.219        & 2.152        & 1.274       &\textbf{18.96} & \textbf{4.620}\\
tallest pinned bubble     & 0.7750 & 1.720        &\textbf{2.642}& 0.8867      & 10.62         & 3.808\\
tallest spreading bubble  & 0.8330 & 2.297        &\textbf{2.213}& 1.126       & 14.48         & 4.223\\
largest stable contact radius& 1.000  &\textbf{3.832}& 0.000     & $\infty$    & 0.000         & 0.000 \\
\hline
\end{tabular}
\end{center}
\end{table}

\section{Conclusions}\label{secConclusions}

\subsection{Answers to the three key questions}

We now return to the three questions posed at the outset. 

\textbf{Answer to Q1.} The detachment volume depends on whether the bubble is pinned or spreading. For pinned bubbles, the detachment volume rises nearly linearly with cavity radius, closely following the Tate prediction $V = 2\pi\lambda^2 r_0$ up to $V = 18.96\,\lambda^3$ with quantitative corrections arising from the non-spherical interface shape under buoyancy (Figure~\ref{figVol}a). For cavities with $3.219<r_0/\lambda<3.832$, the detachment volume is $V<18.96\,\lambda^3$, and for cavities larger than $3.832\,\lambda$ the detachment volume is zero as no stable bubble can exist. For spreading bubbles, the detachment volume increases monotonically with contact angle and saturates at $V = 18.96\,\lambda^3$ as $\phi_0 \to \pi$, below the Fritz prediction (Figure~\ref{figVol}b). Both the pinned and spreading results are confirmed by independent experimental datasets spanning electrolytic, boiling, and carbonated bubbles, as well as pendant drops. 

\textbf{Answer to Q2.} For pinned bubbles with small contact radii ($r_0<\lambda$), the width of the bubble is $\approx2r_0^{1/3}\lambda^{2/3}$. However, for larger contact radii, the bubble is no larger than its contact patch. For spreading bubbles, the width closely approximates the piecewise function $2\times3.832\,\lambda\phi_0/(2\pi)$ for $\phi_0<\pi/2$ and $2\times3.832\,\lambda(\phi_0/\pi)^2$ for $\phi_0>\pi/2$. These empirical findings can be used to control and predict bubble coalescence.

\textbf{Answer to Q3.} The detachment mode is determined by which threshold is reached first during bubble growth (Figure~\ref{figTree}). The behaviour depends strictly on the cavity radius $r_c$ relative to the wetting angle $\phi_w$ and the capillary length $\lambda$, as summarised in Table~\ref{tabDetachment}.

\begin{table}\small
\caption{Bubble detachment modes based on cavity radius $r_c$, wetting angle $\phi_w$, and capillary length $\lambda$.}\label{tabDetachment}
\begin{center}
\begin{tabular}{ccc}
\hline
Cavity radius & Bubble mode & Detachment mode \\
\hline
 $r_c < 3.5\,\lambda\left(\phi_w/\pi\right)^2$ & spreading & straight upwards \\
 $3.5\,\lambda\left(\phi_w/\pi\right)^2 \le r_c < 3.219\,\lambda$ & pinned & straight upwards \\
 $3.219\,\lambda \le r_c < 3.832\,\lambda$ & pinned & sideways \\
 $r_c \ge 3.832\,\lambda$ & no stable bubble & sideways \\
\hline
\end{tabular}
\end{center}
\end{table}

\subsection{Outlook}

In summary, we have computed the maximum stable volume of axisymmetric bubbles on flat surfaces as a function of contact radius and contact angle by numerically solving the Young--Laplace equation. The results unify the pinned and spreading detachment modes within a single framework, recover the classical Tate and Fritz limits, and identify the thresholds at which non-axisymmetric instabilities govern departure. All results scale universally with the capillary length $\lambda$ and also apply to pendant drops. The numerical code is freely available\cite{cannon26bubble} and provides an exact detachment criterion for full electrolyser simulations such as that of\citet{ross26transient}, as well as quantitative guidance for designing electrodes with optimised wettability and cavity geometry to promote early bubble detachment and improve electrolysis efficiency\cite{raman25electrolytic,penas19decoupling}. The results also inform the design of gas-evolving electrodes for space applications. In reduced-gravity environments such as Martian and lunar bases, $\lambda$ increases to $\approx4.6\,\mathrm{mm}$ and $\approx6.9\,\mathrm{mm}$, respectively, compared with $\sim2.8\,\mathrm{mm}$ on Earth, allowing substantially larger stable bubbles.

In other applications, our detachment criteria provide a physical limit for models of environmental gas transport, such as methane seepage. As shown in Table~\ref{tabSig}, the maximum detachment diameter in water is $2r_\mathrm{det}=9.24\,\mathrm{mm}$. This is smaller than the $12.4\,\mathrm{mm}$ diameter required for methane bubbles to reach the atmosphere from deep-water seeps at $\sim230\,\mathrm{m}$ depth\cite{mcginnis06fate}. Thus, individual bubbles detaching from sediment cannot grow sufficiently large to reach the atmosphere from deep water.

Future work could extend the model to heterogeneous surfaces, where contact angle hysteresis and repinning of the contact line can occur.

\section{Data availability}

The data and code that support the findings of this article are openly available on Zenodo at \url{https://doi.org/10.5281/zenodo.20720640} (ref.\citenum{cannon26bubble}).
\renewcommand{\thefootnote}{\alph{footnote}}
\footnotetext{\textsuperscript{a}Center of Applied Space Technology and Microgravity -- ZARM, University of Bremen, 28359 Bremen, Germany.}
\footnotetext{\textsuperscript{b}Faculty of Production Engineering, Mechanical and Process Engineering, and MAPEX Center for Materials and Processes, University of Bremen, 28359 Bremen, Germany.}
\footnotetext{\textsuperscript{c}Leibniz Institute for Materials Engineering -- IWT, University of Bremen, 28359 Bremen, Germany.}
\footnotetext{\textit{$^{\dag}$~E-mail: \href{mailto:ianto.cannon@zarm.uni-bremen.de}{ianto.cannon@zarm.uni-bremen.de}}}

\section{Author Contributions}

I.~Cannon: conceptualisation, software, formal analysis, visualisation, writing -- original draft. S.~Endres: software, writing -- review \& editing. M.~Avila: conceptualisation, methodology, writing -- review \& editing. L.~M\"adler: conceptualisation, methodology, writing -- review \& editing.

\section{Conflicts of interest}

There are no conflicts to declare.

\section{Acknowledgements}

This research was supported by the University of Bremen and the State of Bremen within the Humans on Mars Initiative and partially supported by the Deutsche Forschungsgemeinschaft (DFG, German Research Foundation) under Germany's Excellence Strategy – EXC-3036 The Martian Mindset, project number: 533607631. I.~Cannon gratefully acknowledges financial support from the Alexander von Humboldt Foundation through a Humboldt Research Fellowship for Postdoctoral Researchers.

\bibliography{electrolysis,rsc}

@article{allred21role,
  title = {The Role of Dynamic Wetting Behavior during Bubble Growth and Departure from a Solid Surface},
  author = {Allred, Taylor P. and Weibel, Justin A. and Garimella, Suresh V.},
  year = 2021,
  month = jun,
  journal = {Int. J. Heat Mass Transf.},
  volume = {172},
  pages = {121167},
  issn = {00179310},
  doi = {10.1016/j.ijheatmasstransfer.2021.121167},
  langid = {english}
}

@article{angulo20influence,
  title = {Influence of Bubbles on the Energy Conversion Efficiency of Electrochemical Reactors},
  author = {Angulo, Andrea and van der Linde, Peter and Gardeniers, Han and Modestino, Miguel and Rivas, David Fern{\'a}ndez},
  year = 2020,
  month = mar,
  journal = {Joule},
  volume = {4},
  number = {3},
  pages = {555--579},
  issn = {2542-4785, 2542-4351},
  doi = {10.1016/j.joule.2020.01.005},
  langid = {english}
}

@book{bashforth83attempt,
  title = {An Attempt to Test the Theories of Capillary Action by Comparing the Theoretical and Measured Forms of Drops of Fluid},
  author = {Bashforth, Francis and Adams, John Couch},
  year = 1883,
  publisher = {Cambridge University Press},
  collaborator = {{University Of California Libraries}},
  langid = {english},
  lccn = {ucb:GLAD-17060080}
}

@article{bonjour00experimental,
  title = {Experimental Study of the Coalescence Phenomenon during Nucleate Pool Boiling},
  author = {Bonjour, Jocelyn and Clausse, Marc and Lallemand, Monique},
  year = 2000,
  month = feb,
  journal = {Exp. Therm. Fluid Sci.},
  volume = {20},
  number = {3},
  pages = {180--187},
  issn = {0894-1777},
  doi = {10.1016/S0894-1777(99)00044-8},
  langid = {english}
}

@article{bosch32development,
  title = {The Development of the Chemical High Pressure Method during the Establishment of the New Ammonia Industry},
  author = {Bosch, Carl},
  year = 1932,
  month = may,
  journal = {Nobel Lect.},
  pages = {197--241},
  address = {Stockholm, Sweden},
  langid = {english}
}

@article{brinkert22fundamentals,
  title = {Fundamentals and Future Applications of Electrochemical Energy Conversion in Space},
  author = {Brinkert, Katharina and Mandin, Philippe},
  year = 2022,
  month = nov,
  journal = {npj Microgravity},
  volume = {8},
  number = {1},
  pages = {52},
  publisher = {Nature Publishing Group},
  issn = {2373-8065},
  doi = {10.1038/s41526-022-00242-3},
  copyright = {2022 The Author(s)},
  langid = {english}
}

@misc{cannon26bubble,
  title = {Bubble},
  author = {Cannon, Ianto},
  year = 2026,
  month = jun,
  doi = {10.5281/zenodo.20720640},
  howpublished = {Zenodo}
}

@article{chesters77analytical,
  title = {An Analytical Solution for the Profile and Volume of a Small Drop or Bubble Symmetrical about a Vertical Axis},
  author = {Chesters, A. K.},
  year = 1977,
  month = aug,
  journal = {J. Fluid Mech.},
  volume = {81},
  number = {4},
  pages = {609--624},
  issn = {0022-1120, 1469-7645},
  doi = {10.1017/S0022112077002250},
  copyright = {https://www.cambridge.org/core/terms},
  langid = {english}
}

@article{chesters78modes,
  title = {Modes of Bubble Growth in the Slow-Formation Regime of Nucleate Pool Boiling},
  author = {Chesters, A. K.},
  year = 1978,
  month = aug,
  journal = {Int. J. Multiph. Flow},
  volume = {4},
  number = {3},
  pages = {279--302},
  issn = {0301-9322},
  doi = {10.1016/0301-9322(78)90003-4},
  langid = {english}
}

@article{chi-yeh65mechanism,
  title = {The Mechanism of Heat Transfer in Nucleate Pool Boiling---Part {{I}}},
  author = {{Chi-Yeh}, Han and Griffith, Peter},
  year = 1965,
  month = jun,
  journal = {Int. J. Heat Mass Transf.},
  volume = {8},
  number = {6},
  pages = {887--904},
  issn = {00179310},
  doi = {10.1016/0017-9310(65)90073-6},
  copyright = {https://www.elsevier.com/tdm/userlicense/1.0/},
  langid = {english}
}

@book{degennes04capillarity,
  title = {Capillarity and Wetting Phenomena},
  author = {De Gennes, Pierre-Gilles and {Brochard-Wyart}, Fran{\c c}oise and Qu{\'e}r{\'e}, David},
  year = 2004,
  publisher = {Springer New York},
  address = {New York, NY},
  doi = {10.1007/978-0-387-21656-0},
  copyright = {http://www.springer.com/tdm},
  isbn = {978-1-4419-1833-8 978-0-387-21656-0},
  langid = {english}
}

@article{degennes85wetting,
  title = {Wetting: Statics and Dynamics},
  shorttitle = {Wetting},
  author = {{de Gennes}, P. G.},
  year = 1985,
  month = jul,
  journal = {Rev. Mod. Phys.},
  volume = {57},
  number = {3},
  pages = {827--863},
  doi = {10.1103/RevModPhys.57.827},
  langid = {english}
}

@article{demirkir24life,
  title = {Life beyond Fritz: On the Detachment of Electrolytic Bubbles},
  shorttitle = {Life beyond Fritz},
  author = {Demirk{\i}r, {\c C}ayan and Wood, Jeffery A. and Lohse, Detlef and Krug, Dominik},
  year = 2024,
  month = oct,
  journal = {Langmuir},
  volume = {40},
  number = {39},
  pages = {20474--20484},
  issn = {0743-7463},
  doi = {10.1021/acs.langmuir.4c01963},
  langid = {english}
}

@article{demirkir25jump,
  title = {To Jump or Not to Jump: Adhesion and Viscous Dissipation Dictate the Detachment of Coalescing Wall-Attached Bubbles},
  shorttitle = {To Jump or Not to Jump},
  author = {Demirk{\i}r, {\c C}ayan and Yang, Rui and Bashkatov, Aleksandr and Sanjay, Vatsal and Lohse, Detlef and Krug, Dominik},
  year = 2025,
  month = dec,
  journal = {Phys. Rev. Fluids},
  volume = {10},
  number = {12},
  pages = {123602},
  publisher = {American Physical Society},
  doi = {10.1103/nq3w-57gg},
  langid = {english}
}

@article{franchi20hydrogen,
  title = {Hydrogen Production via Steam Reforming: A Critical Analysis of {{MR}} and {{RMM}} Technologies},
  shorttitle = {Hydrogen Production via Steam Reforming},
  author = {Franchi, Giovanni and Capocelli, Mauro and Falco, Marcello De and Piemonte, Vincenzo and Barba, Diego},
  year = 2020,
  month = jan,
  journal = {Membranes},
  volume = {10},
  number = {1},
  pages = {10},
  publisher = {Multidisciplinary Digital Publishing Institute},
  issn = {2077-0375},
  doi = {10.3390/membranes10010010},
  copyright = {http://creativecommons.org/licenses/by/3.0/},
  langid = {english}
}

@article{fritz35berechnung,
  title = {Berechnung Des Maximalvolumes von Dampfblasen},
  author = {Fritz, W.},
  year = 1935,
  journal = {Phys. Zeitschr},
  volume = {36},
  pages = {379--384},
  langid = {english}
}

@article{gerlach05quasistatic,
  title = {Quasi-Static Bubble Formation on Submerged Orifices},
  author = {Gerlach, D. and Biswas, G. and Durst, F. and Kolobaric, V.},
  year = 2005,
  month = jan,
  journal = {Int. J. Heat Mass Transf.},
  volume = {48},
  number = {2},
  pages = {425--438},
  publisher = {Elsevier BV},
  issn = {0017-9310},
  doi = {10.1016/j.ijheatmasstransfer.2004.09.002},
  copyright = {https://www.elsevier.com/tdm/userlicense/1.0/},
  langid = {english}
}

@article{gunde01measurement,
  title = {Measurement of the Surface and Interfacial Tension from Maximum Volume of a Pendant Drop},
  author = {Gunde, Rok and Kumar, Arun and {Lehnert-Batar}, Stefanie and M{\"a}der, Roland and Windhab, Erich J.},
  year = 2001,
  month = dec,
  journal = {J. Colloid Interface Sci.},
  volume = {244},
  number = {1},
  pages = {113--122},
  issn = {0021-9797},
  doi = {10.1006/jcis.2001.7916},
  langid = {english}
}

@article{huang26effects,
  title = {Effects of Surface Wettability on Bubble Dynamics and Induced Liquid Flow: Finite-Difference Analysis of Two-Phase Particle Image Velocimetry},
  shorttitle = {Effects of Surface Wettability on Bubble Dynamics and Induced Liquid Flow},
  author = {Huang, Jianxun and Li, Ri},
  year = 2026,
  month = feb,
  journal = {Phys. Rev. Fluids},
  volume = {11},
  number = {2},
  pages = {023603},
  issn = {2469-990X},
  doi = {10.1103/jvxz-8mzv},
  langid = {english}
}

@article{kullmann23role,
  title = {The Role of Hydrogen for the Defossilization of the German Chemical Industry},
  author = {Kullmann, Felix and Lin{\ss}en, Jochen and Stolten, Detlef},
  year = 2023,
  month = dec,
  journal = {Int. J. Hydrog. Energy},
  volume = {48},
  number = {99},
  pages = {38936--38952},
  issn = {03603199},
  doi = {10.1016/j.ijhydene.2023.04.191},
  langid = {english}
}

@book{laplace98traite,
  title = {Trait\'e de M\'ecanique C\'eleste},
  author = {Laplace, Pierre Simon},
  year = 1798,
  publisher = {Courcier},
  address = {Paris},
  langid = {english},
  lccn = {39088005644729}
}

@article{lesage13experimental,
  title = {Experimental and Numerical Analysis of Quasi-Static Bubble Size and Shape Characteristics at Detachment},
  author = {Lesage, Fr{\'e}d{\'e}ric J. and Marois, Francis},
  year = 2013,
  month = sep,
  journal = {Int. J. Heat Mass Transf.},
  volume = {64},
  pages = {53--69},
  publisher = {Elsevier BV},
  issn = {0017-9310},
  doi = {10.1016/j.ijheatmasstransfer.2013.04.019},
  copyright = {https://www.elsevier.com/tdm/userlicense/1.0/},
  langid = {english}
}

@article{matsushima06water,
  title = {Water Electrolysis under Microgravity},
  author = {Matsushima, H. and Fukunaka, Y. and Kuribayashi, K.},
  year = 2006,
  month = may,
  journal = {Electrochimica Acta},
  volume = {51},
  number = {20},
  pages = {4190--4198},
  issn = {00134686},
  doi = {10.1016/j.electacta.2005.11.046},
  copyright = {https://www.elsevier.com/tdm/userlicense/1.0/},
  langid = {english}
}

@article{mcginnis06fate,
  title = {Fate of Rising Methane Bubbles in Stratified Waters: How Much Methane Reaches the Atmosphere?},
  shorttitle = {Fate of Rising Methane Bubbles in Stratified Waters},
  author = {McGinnis, D. F. and Greinert, J. and Artemov, Y. and Beaubien, S. E. and W{\"u}est, A.},
  year = 2006,
  journal = {J. Geophys. Res. Oceans},
  volume = {111},
  number = {C9},
  pages = {C09007},
  issn = {2156-2202},
  doi = {10.1029/2005JC003183},
  langid = {english}
}

@article{michael76equilibrium,
  title = {The Equilibrium and Stability of Axisymmetric Pendent Drops},
  author = {Michael, D. H. and Williams, P. G.},
  year = 1976,
  journal = {Proc. R. Soc. Lond. Ser. Math. Phys. Sci.},
  volume = {351},
  number = {1664},
  eprint = {79312},
  eprinttype = {jstor},
  pages = {117--127},
  issn = {0080-4630},
  doi = {10.1098/rspa.1976.0132},
  langid = {english}
}

@article{mori01bubble,
  title = {Bubble Departure from Cavities},
  author = {Mori, Brian K and Baines, W.Douglas},
  year = 2001,
  month = feb,
  journal = {Int. J. Heat Mass Transf.},
  volume = {44},
  number = {4},
  pages = {771--783},
  issn = {0017-9310},
  doi = {10.1016/s0017-9310(00)00133-2},
  copyright = {https://www.elsevier.com/tdm/userlicense/1.0/},
  langid = {english}
}

@article{mukherjee07numerical,
  title = {Numerical Study of Single Bubbles with Dynamic Contact Angle during Nucleate Pool Boiling},
  author = {Mukherjee, Abhijit and Kandlikar, Satish G.},
  year = 2007,
  month = jan,
  journal = {Int. J. Heat Mass Transf.},
  volume = {50},
  number = {1-2},
  pages = {127--138},
  issn = {00179310},
  doi = {10.1016/j.ijheatmasstransfer.2006.06.037},
  copyright = {https://www.elsevier.com/tdm/userlicense/1.0/},
  langid = {english}
}

@article{penas19decoupling,
  title = {Decoupling Gas Evolution from Water-Splitting Electrodes},
  author = {Pe{\~n}as, Pablo and van der Linde, Peter and Vijselaar, Wouter and van der Meer, Devaraj and Lohse, Detlef and Huskens, Jurriaan and Gardeniers, Han and Modestino, Miguel A. and Rivas, David Fern{\'a}ndez},
  year = 2019,
  month = oct,
  journal = {J. Electrochem. Soc.},
  volume = {166},
  number = {15},
  pages = {H769},
  publisher = {IOP Publishing},
  issn = {1945-7111},
  doi = {10.1149/2.1381914jes},
  langid = {english}
}

@article{phan09surface,
  title = {Surface Wettability Control by Nanocoating: The Effects on Pool Boiling Heat Transfer and Nucleation Mechanism},
  shorttitle = {Surface Wettability Control by Nanocoating},
  author = {Phan, Hai Trieu and Caney, Nadia and Marty, Philippe and Colasson, St{\'e}phane and Gavillet, J{\'e}r{\^o}me},
  year = 2009,
  month = nov,
  journal = {Int. J. Heat Mass Transf.},
  volume = {52},
  number = {23-24},
  pages = {5459--5471},
  issn = {00179310},
  doi = {10.1016/j.ijheatmasstransfer.2009.06.032},
  copyright = {https://www.elsevier.com/tdm/userlicense/1.0/},
  langid = {english}
}

@book{plateau73statique,
  title = {Statique Exp\'erimentale et Th\'eorique Des Liquides Soumis Aux Seules Forces Mol\'eculaires},
  author = {Plateau, Joseph Antoine Ferdinand},
  year = 1873,
  publisher = {Paris, Gauthier-Villars},
  collaborator = {{Unknown Library}},
  langid = {english}
}

@article{raman25electrolytic,
  title = {Electrolytic Bubble Coalescence on Hydrophobic Cavity Arrays Determines Departure Radius and Lowers Electrolyte Supersaturation},
  author = {Raman, Akash and Schlautmann, Stefan and Gardeniers, Han and Rivas, David Fern{\'a}ndez},
  year = 2025,
  month = nov,
  journal = {Small},
  volume = {21},
  number = {44},
  pages = {e05728},
  issn = {1613-6810, 1613-6829},
  doi = {10.1002/smll.202505728},
  langid = {english}
}

@article{ross25impact,
  title = {Impact of Gas Bubble Evolution Dynamics on Electrochemical Reaction Overpotentials in Water Electrolyser Systems},
  author = {Ross, Byron and Haussener, Sophia and Brinkert, Katharina},
  year = 2025,
  month = mar,
  journal = {J. Phys. Chem. C},
  volume = {129},
  number = {9},
  pages = {4383--4397},
  issn = {1932-7447, 1932-7455},
  doi = {10.1021/acs.jpcc.5c00220},
  copyright = {https://creativecommons.org/licenses/by/4.0/},
  langid = {english}
}

@article{ross26transient,
  title = {Transient Simulation of Gas Bubble Evolution and Overpotential Dynamics for the Hydrogen Evolution Reaction},
  author = {Ross, Byron and Skidmore, Kayleigh and Haussener, Sophia and Brinkert, Katharina},
  year = 2026,
  month = jan,
  journal = {ACS Electrochem.},
  volume = {2},
  number = {1},
  pages = {113--123},
  publisher = {American Chemical Society},
  doi = {10.1021/acselectrochem.5c00291},
  langid = {english}
}

@article{sasetty23stability,
  title = {Stability and Critical Volume of a Suspended Pendant Drop in Air via Experiments and Eigenvalue Analysis},
  author = {Sasetty, Sravya and Ward, Thomas},
  year = 2023,
  month = jun,
  journal = {Colloids Surf. Physicochem. Eng. Asp.},
  volume = {666},
  pages = {131346},
  issn = {09277757},
  doi = {10.1016/j.colsurfa.2023.131346},
  langid = {english}
}

@article{steffen18trajectories,
  title = {Trajectories of the {{Earth}} System in the Anthropocene},
  author = {Steffen, Will and Rockstr{\"o}m, Johan and Richardson, Katherine and Lenton, Timothy M. and Folke, Carl and Liverman, Diana and Summerhayes, Colin P. and Barnosky, Anthony D. and Cornell, Sarah E. and Crucifix, Michel and Donges, Jonathan F. and Fetzer, Ingo and Lade, Steven J. and Scheffer, Marten and Winkelmann, Ricarda and Schellnhuber, Hans Joachim},
  year = 2018,
  month = aug,
  journal = {Proc. Natl. Acad. Sci.},
  volume = {115},
  number = {33},
  pages = {8252--8259},
  publisher = {Proceedings of the National Academy of Sciences},
  doi = {10.1073/pnas.1810141115},
  langid = {english}
}

@article{sumesh10possible,
  title = {The Possible Equilibrium Shapes of Static Pendant Drops},
  author = {Sumesh, P. T. and Govindarajan, Rama},
  year = 2010,
  month = oct,
  journal = {J. Chem. Phys.},
  volume = {133},
  number = {14},
  pages = {144707},
  issn = {0021-9606},
  doi = {10.1063/1.3494041},
  langid = {english}
}

@article{tate64magnitude,
  title = {On the Magnitude of a Drop of Liquid Formed under Different Circumstances},
  author = {Tate, T.},
  year = 1864,
  month = mar,
  journal = {Lond. Edinb. Dublin Philos. Mag. J. Sci.},
  volume = {27},
  number = {181},
  pages = {176--180},
  issn = {1941-5982, 1941-5990},
  doi = {10.1080/14786446408643645},
  langid = {english}
}

@article{vogt89problem,
  title = {The Problem of the Departure Diameter of Bubbles at Gas-Evolving Electrodes},
  author = {Vogt, H.},
  year = 1989,
  month = oct,
  journal = {Electrochimica Acta},
  volume = {34},
  number = {10},
  pages = {1429--1432},
  issn = {00134686},
  doi = {10.1016/0013-4686(89)87183-X},
  copyright = {https://www.elsevier.com/tdm/userlicense/1.0/},
  langid = {english}
}

@article{weber19global,
  title = {Global Ocean Methane Emissions Dominated by Shallow Coastal Waters},
  author = {Weber, Thomas and Wiseman, Nicola A. and Kock, Annette},
  year = 2019,
  month = oct,
  journal = {Nat. Commun.},
  volume = {10},
  number = {1},
  pages = {4584},
  publisher = {Nature Publishing Group},
  issn = {2041-1723},
  doi = {10.1038/s41467-019-12541-7},
  copyright = {2019 The Author(s)},
  langid = {english}
}

@Control{rsc-control,
  ctrl-use-doi-all = {yes},
  ctrl-use-title = {yes}
}

\end{document}